# Event-driven Two-stage Solution to Non-intrusive Load Monitoring

Lei Yan[a], Wei Tian[a], Jiayu Han[a], Zuyi Li[a*]

[a]Illinois Institute of Technology, Electrical and Computer Engineering Department, Chicago, IL 60616 USA
*Corresponding Author: Zuyi Li (e-mail: lizu@iit.edu).

HIGHLIGHTS

- An event-driven Factorial Hidden Markov model (eFHMM) for multiple appliances is proposed to greatly reduce the computational complexity.
- The eFHMM is solved in two stages using transient signatures and steady-state signatures, termed as eFHMM-TS, to ensure accurate load disaggregation.
- An edge-cloud framework is designed for implementing eFHMM-TS to realize real-world application in real time.

ARTICLE INFO

ABSTRACT

Existing methods of non-intrusive load monitoring (NILM) in literatures generally suffer from high computational complexity and/or low accuracy in identifying working household appliances. This paper proposes an event-driven Factorial Hidden Markov model (eFHMM) for multiple appliances with multiple states in a household, aiming for low computational complexity and high load disaggregation accuracy. The proposed eFHMM decreases the computational complexity to be linear to the event number, which ensures online load disaggregation. Furthermore, the eFHMM is solved in two stages, where the first stage identifies state-changing appliance using transient signatures and the second stage confirms the inferred states using steady-state signatures. The combination of transient and steady-state signatures, which are extracted from transient and steady periods segmented by detected events, enhances the uniqueness of each state transition and associated appliances, which ensures accurate load disaggregation. The event-driven two-stage NILM solution, termed as eFHMM-TS, is naturally fit into an edge-cloud framework, which makes possible the real-world application of NILM. The proposed eFHMM-TS method is validated on the LIFTED and synD datasets. Results demonstrate that the eFHMM-TS method outperforms other methods and can be applied in practice.

## 1. Introduction

Residential and commercial buildings consume about 73.5% of produced electricity in 2019 in USA [1]. Research shows that real-time feedback of detailed power information of individual appliances can efficiently save up to 12% of total power consumption [2]. Nonintrusive load monitoring (NILM), initiated by Hart [3], provides a cost-effective solution to identifying appliances and their power usage from composite loads.

As the world's most urgent mission by 2050 is carbon neutrality, most countries have committed to achieve net zero emission [4]. This will lead to high penetration of renewable energy such as solar and wind energy that may not generate enough electricity to meet peak demand at night. NILM enables electric utility to send real-time power usage to consumers inspiring their green behaviors and to directly control load by turning targeted appliances such as air conditioners and water heaters off during peak periods.

Most NILM methods proposed in literatures are mainly based on graph signal processing (GSP) [5, 6], deep neural network (DNN) and Hidden Markov Model (HMM). Neural-NILM [7] proposes three neural network architectures for energy disaggregation, being the first of its kind. Subsequent works further develop neural network based NILM to enhance the performance. Ref. [8] applies gated recurrent unit (GRU) based on recurrent neural network (RNN) to classify the activations of target appliances. A convolutional neural network (CNN) based method classifies the switching on events in [9]. Ref. [10] modifies seq2seq learning to seq2point learning in CNN structure to decrease computational cost. Their results indicate that these models struggle to achieve high accuracy for multi-state appliances with repeatable switching patterns. It is further developed to apply the trained model to other households using transfer learning [11]. Ref. [12] uses long short-term memory (LSTM) to classify appliances based on denoising autoencoder. As DNN-based NILM methods require a large training dataset to tune parameters with higher computational complexity than other methods, it deserves further careful consideration and analysis.

HMM is another major type of solution to NILM. It is based on Markov chain that describes transition probability among multiple different states. A Markov chain can be used to model the transitions among different states of an appliance. Appliance states of interest are hidden so they cannot be observed directly. HMM aims to infer an appliance's latent states from observed measurements, which makes HMM naturally suited for modeling an appliance. Factorial HMM (FHMM) is an extension of HMM, which includes multiple independent Markov chains of hidden states, and each chain represents one appliance. A household with many appliances can be modeled using FHMM.

Recently, many variants of HMM have been reported to better model appliances in a household. Ref. [13] uses an iterative hidden model to separate individual appliances from the aggregated load iteratively with tuned appliance models. Ref. [14] adopts two FHMM variants, i.e., additive and difference FHMM, to model energy disaggregation problem. It develops a convex formulation of approximation inference instead of exact inference to separate appliances. Load Disaggregation is performed using an alternative formulation of additive Factorial Approximate Maximum a Posteriori (AFAMAP) [15]. An Adaptive Density Peak Clustering-Factorial Hidden Markov model (ADP-FHMM) reduces the dependence of prior information and automatically determine the working appliances based on power consumption [16]. A hybrid signature-based iterative disaggregation based on the combination of Factorial Hidden Markov model is proposed to improve performance when multiple appliances are operated simultaneously [17]. A segmented integer quadratic constraint programming (SIQCP) is proposed to model home appliances as HMMs and identify appliances [18]. Ref. [19] propose a conditional factorial hidden semi-Markov model (CFHSMM) integrating additional time-related features to disaggregate load using maximum likelihood estimation.

These variants lead to further development of HMM-based NILM, but some associated problems have not been completely solved. Some of these methods struggle to achieve a high accuracy as in [14]. The computational cost of some methods is too high for the real-time application due to their complex framework and massive variable matrix such as [14, 18]. This issue is partially addressed in [20] by modeling a single household with a gigantic but sparse super-state HMM and using the solver of sparse Viterbi algorithm. However, scalability is one of its major drawbacks that limits its application in practice. Besides, some methods only work well for type I appliances with only ON and OFF states such as kettle, toaster, and table lamp. Ref. [21] proposes a load disaggregation method specifically for type I appliance based on subtractive clustering and maximum likelihood classifier that limits its real-world applications. Type II appliances are the most common ones in residential households, which have multiple states such as washing machine and hair dryer. As defined in [22], there are also type III and type IV appliances. Type III appliances continuously change power consumption such as electric drill, which is infrequently used, and type IV appliances always consume small amount of power which are not of interest in the context of NILM.

To better evaluate performance of different NILM methods, NILM-TK [23] is proposed as an open-source toolkit to enable comparison among different NILM methods in a reproducible way. It gives a complete pipeline from multiple datasets to evaluation metrics. It also includes several disaggregation benchmark algorithms such as combinatorial optimization (CO) and factorial hidden Markov model (FHMM).

Event-based NILM is an alternative to solve the issues in existing HMM-based methods. Ref. [21] is an event-based load disaggregation method, but it only works for type I appliances as in [24] that proposes an event matching algorithm for ON/OFF events. Ref. [25] models each appliance with triangle and rectangle signatures, and it classifies appliances using range of power change that may misidentify appliances with similar signatures. An unsupervised event-based method in [26] is designed using additive FHMM, but it does not consider the transient signatures to infer associated appliances. Transient signatures have the potential of better identifying appliances, thus deserving further careful study.

As the change of appliances' working state is always accompanied by the occurrence of an event, it is natural to reformulate the conventional time-driven FHMM to an event-driven FHMM, termed as eFHMM in this paper. In fact, it is not necessary to frequently infer appliances' states in steady periods. Doing that otherwise might be the main reason why most NILM methods based on HMM take a long even exponential time to obtain solution. In the proposed eFHMM framework, the inference times are as limited as the appliances' state-changing times. Furthermore, an event-driven two-stage NILM solution, termed as eFHMM-TS, is proposed to improve the accuracy of load disaggregation, where transient load signatures are adopted to infer which appliance is changing state in the first stage and steady-state load signatures are used to confirm the obtained results by comparing composite total and estimated total load in the second stage.

In general, an event-driven NILM solution requires data sampling rate larger than 1 Hz to guarantee the usability of transient signatures. It puts huge burden on communication network bandwidth to make associated data available to the remote server. To solve this problem, eFHMM-TS can be configured naturally in an edge-cloud framework by dividing the whole NILM task into two parts that are completed in edge and cloud sides to further release the burden of network bandwidth and local computing system in a reliable way.

The contributions of this paper are three-fold.
- First, this paper proposes an event-driven Factorial Hidden Markov Model (eFHMM) to model the appliances in a household. The proposed eFHMM enables online load disaggregation by decreasing inference times to be linear to detected event number thus greatly reducing computational complexity.
- Second, this paper specifically designs a two-stage NILM solution to eFHMM, with the first stage inferring changing states using transient signatures and the second stage confirming the estimated states with steady-state signatures. The combination of transient and steady-state signatures ensures the uniqueness of transition states thus leading to a higher load disaggregation accuracy.
- Third, this paper proposes the implementation of eFHMM-TS in an edge-cloud framework, where the data-intensive event detection is done at the edge side and the model-intensive load disaggregation is completed at the cloud side. Such an implementation can minimize the requirements on computing capacity, storage space, and available communication bandwidth for resource-poor IoT devices, and along with the inherent protection of data privacy, making it possible to apply NILM in real world.

The rest of this paper is organized as follows. The eFHMM is introduced in Section 2. Section 3 presents the two-stage solution to eFHMM. Section 4 integrates the eFHMM-TS into the edge-cloud framework. Section 5 presents and discusses experimental results. Finally, Section 6 concludes this paper.

## 2. Event-driven Factorial HMM

Modeling of individual appliances and the combination of appliances in a household is the core of NILM. This section will briefly introduce the motivation of event-driven NILM and the modeling methodology of sparse HMM and eFHMM.

### 2.1 Motivation of event-driven NILM

The objective of NILM is to infer appliances' states from aggregated load. One fact to be noted is that each appliance changes states only a limited number of times during a given period of time. For example, a toaster is generally operated only several times a week. So, it is not necessary to perform point-to-point inference, which might be the main reason why some methods suffer from high computational cost. In event-driven NILM, the states are only inferred when an event occurs. As there are limited appliances with limited events in a household, the inference times are limited as well thus greatly reducing the computational complexity. A detailed complexity analysis can be found in Section 3.2.

Beyond decreasing computational cost, an event-driven approach can extract transient signatures from transient processes when events occur, which can play a decisive role in accurately identifying individual appliances. Since appliances' power signatures inherit from their electronic components, appliances with different components display different transient waveshapes and power values as shown in Fig. 1.

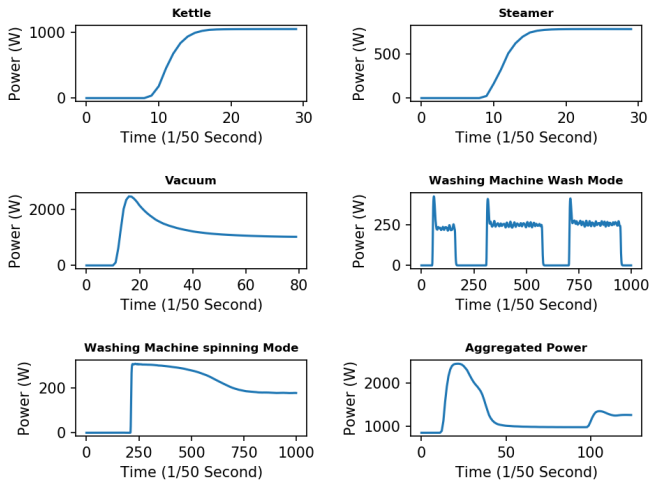

Fig. 1 Appliance power consumption during transients.

Residential appliances are made of basic components such as heating resistor (HR), motor, and electronic circuit (EC). HR determines the active power based on 'Joule's Law' with a relatively stable value as the change of HR is tiny during working state. It will quickly reach steady-state period once it starts working. Motor converts electrical energy to mechanical energy through electromagnetic induction that would cause transient power to reach a peak value then decrease to a steady-state value. The transient process of motor appliances usually takes a longer time than that of HR appliances. As shown in Fig. 1, kettle and steamer quickly reach the steady state without a transient peak; however, vacuum and washing machine (wash and spin mode) experience a more complex transient process which includes a transient peak and a gradual decrease that lasts a long time. Appliances with EC usually generate small amount of power with high fluctuation. Their transient processes are comparatively more complex.

Therefore, different kinds of appliances generate different transient processes. Event-driven NILM makes use of these unique characteristics and enjoys advantages such as lower computational complexity and more distinctive load signatures. If these characteristics are fully exploited, a highly accurate and real-time NILM can be achieved.

### 2.2 Event detection and signature extraction

Event detection is the first step in event-driven NILM and it can provide decisive information to identify appliances. The detected event can divide the overall electrical measurements into transient and steady periods from which load signatures including both transient signatures and steady-state signatures are extracted.

Nearly all existing methods [27-28] tend to detect a change-point rather than the overall transient period. The event detection method used in this paper is from our previous work [29]. It enables online event detection and adopts dynamic time-window approach to deal with challenges such as long transition, high fluctuation, and near simultaneity.

Load signatures are key information to distinguish different appliances. As introduced before, different appliances are made of different electrical components that determine their unique transient signatures. It is also important to integrate transient signatures into the mathematical formulation of NILM. Therefore, transient signatures should be quantified and modeled as a mathematical distribution such as the Gaussian distribution. To the best of our knowledge, it is the first time to quantitatively analyze the transient signatures.

The outputs from our event detection method [29] include starting point, transient spike point, and ending point as shown in Fig. 2. Based on these outputs, transient signatures can be extracted such as difference between transient spike value and previous steady-state mean value (DTS), and difference between steady-state mean of datapoints within the immediate next window after the transient and that within the previous one window (DSP). Steady-state signatures such as steady-state power (SSP) are extracted from steady periods.

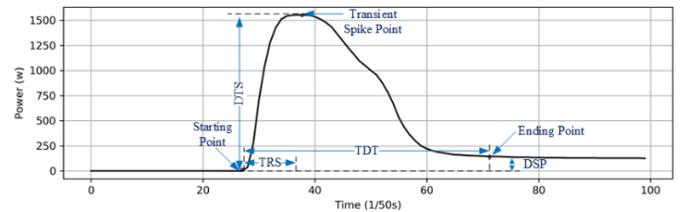

Fig. 2 Illustration of load signatures in refrigerator transition process.

Compared with other works, this paper transforms transient signatures analysis from a qualitive nature to a quantitative nature. It does not just use the transient waveshape to classify whether appliances have transient processes or not. Once an event is detected, transient and steady-state signatures can be extracted. The whole process can be summarized as follows.
- Divide overall electrical measurements into transient and steady periods once events are detected using method in [29].
- Use the modified mean-shift method to determine the state

number of each appliance, which helps cluster steady periods thus extract steady-state signatures such as SSP as statistical parameters.
- Label transitions between two steady periods segmented by events to extract transient signatures such as DTS and DSP as statistical parameters.

Mean-shift clustering was initially proposed in [30] and further developed in [31]. It can automatically determine the number of states without predefining the number. Residential appliances have the common characteristics that one appliance has limited number of states and the powers of different states vary greatly. Specifically, the difference between two high-power states tends to be big. For example, the power values of a refrigerator in LIFTED [32] with different states are [0.52, 42, 121, 156], while the power values of a hair dryer are [0.23, 178, 680, 1230]. The larger the power value of a state, the larger is its difference with the power value of another state. These inherent properties can help design faster mean-shift algorithms effectively by 1) setting different bandwidths for different power levels to determine the state number and the associated power closer to truth; 2) removing duplicate or close centroids in each iteration to accelerate algorithm convergency.

*2.3 Model individual appliances as sparse HMMs*

Assume that one appliance has $K$ states $S = \{1,2,...,K\}$ and its normal operation switches among the $K$ states. For a period of measurements with $T$ time steps, sequence of measurements and operating states can be represented as $x = \{x_1, x_2, ..., x_T\}$ and $z = \{z_1, z_2, ..., z_T\}$. These are the basic information conveyed by the basic Markov chains and HMM. It is true that a state can transfer to itself in the basic Markov chains when using point-to-point inference in NILM. However, it will increase inference times, which limits the application of NILM in practice.

These problems can be completely solved using the event-driven approach. The matrix associated with Markov chains and HMM should be redefined to meet requirement of event-driven NILM. The state transition diagram of an appliance with three states ($K$=3) is illustrated in Fig. 3, where only transitions between states are considered. The emission power of a steady state is defined as $p(x_t|z_t, \emptyset)$, where $\emptyset$ is the emission vector, governing the probability distribution of steady-state power.

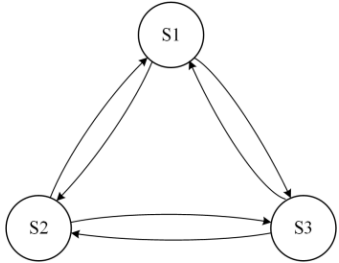

Fig. 3 State transition diagram of an appliance with three states.

Apart from the parameters in the basic HMM, the event-driven NILM proposed in this paper includes two transient signatures, i.e., DTS and DSP, which together can represent a more unique transition process. Their probability distributions are denoted by $B$ and $C$, respectively.

Generally, a complete event-driven HMM is defined by four parameters:

*Transition matrix* $A \in R^{K \times K}$, whose element $A_{i,j}$ represents the transition probability from state $i$ at $t$ - 1 to state $j$ at $t$.

$$A_{i,j} = p(z_{t-1} = i|z_t = j) \quad (1)$$

*DTS matrix* $B = \{B_{1,2}, B_{1,3}, ..., B_{K-1,K}\}$, whose element $B_{i,j}$ is modeled as Gaussian Distribution as in Eq. (2)

$$B_{i,j} \sim N(\mu_{P_{i,j}}, \sigma_{P_{i,j}}) \quad (2)$$

*DSP matrix* $C = \{C_{1,2}, C_{1,3}, ..., C_{K-1,K}\}$, whose element $C_{i,j}$ is modeled as Gaussian Distribution as in Eq. (3).

$$C_{i,j} \sim N(\mu_{S_{i,j}}, \sigma_{S_{i,j}}) \quad (3)$$

*SSP emission vector* $\phi = \{\phi_1, \phi_2, ..., \phi_K\}$, whose element $\phi_i$ is modeled as Gaussian Distribution as in Eq. (4).

$$\phi_i \sim N(\mu_i, \sigma_i) \quad (4)$$

After detecting event and determining state number of each appliance, each period of measurements is clustered to certain state of certain appliance. Correspondingly, mean values of different steady periods belonging to the same state is in the same cluster $c_i$, which is used to represent the corresponding period.

Let $\mathbf{1}_{c_i}(u)$ denotes $u$ belongs to cluster $c_i$ and $\mathbf{1}_{i,j}(u_n)$ denote $\{u_n \in c_j | u_{n-1} \in c_i\}$. $N_{i,j}$ denotes counts of transition process from state $i$ to state $j$.

$$\mathbf{1}_{c_i}(u) := \begin{cases} 1 \text{ if } u \in c_i \\ 0 \text{ if } u \notin c_i \end{cases} \quad (5)$$

$$\mathbf{1}_{i,j}(u_n) := \begin{cases} 1 \text{ if } \mathbf{1}_{c_i}(u_{n-1}) = 1 \& \mathbf{1}_{c_j}(u_n) = 1 \\ 0 \text{ otherwise} \end{cases} \quad (6)$$

$$N_{i,j} = \sum_{n=2:N} \mathbf{1}_{i,j}(u_n) \quad (7)$$

$$A_{i,j} = \frac{N_{i,j}}{N-1} \quad (8)$$

Let $\mathbf{1}_{p_{i,j}}(DTS_n)$ denotes DTS of the $n^{th}$ transient process from state $i$ to state $j$. $s_{i,j}$ is the cluster for $\mathbf{1}_{s_{i,j}}(DTS_n) = 1$.

$$\mathbf{1}_{p_{i,j}}(DTS_n) := \begin{cases} 1 \text{ if } \mathbf{1}_{c_i}(u_{n-1}) = 1 \& \mathbf{1}_{c_j}(u_n) = 1 \\ 0 \text{ otherwise} \end{cases} \quad (9)$$

$$\mu_{p_{i,j}} = \frac{1}{N_{i,j}} \sum_{DTS \in p_{i,j}} DTS \quad (10)$$

$$\sigma_{p_{i,j}} = sqrt\left(\frac{1}{N_{i,j}} \sum_{DTS \in p_{i,j}} \left(DTS - \mu_{p_{i,j}}\right)^2\right) \quad (11)$$

Let $\mathbf{1}_{s_{i,j}}(DSP_n)$ denote $DSP$ of the $n^{th}$ transient process from state $i$ to state $j$. $s_{i,j}$ is the cluster for $\mathbf{1}_{s_{i,j}}(DSP_n) = 1$.

$$\mathbf{1}_{s_{i,j}}(DSP_n) := \begin{cases} 1 \text{ if } \mathbf{1}_{c_i}(u_{n-1}) = 1 \& \mathbf{1}_{c_j}(u_n) = 1 \\ 0 \text{ otherwise} \end{cases} \quad (12)$$

$$\mu_{s_{i,j}} = \frac{1}{N_{i,j}} \sum_{DSP \in s_{i,j}} DSP \quad (13)$$

$$\sigma_{s_{i,j}} = sqrt\left(\frac{1}{N_{i,j}} \sum_{DSP \in s_{i,j}} \left(DSP - \mu_{s_{i,j}}\right)^2\right) \quad (14)$$

Let $\mathbf{1}_{c_i}(x_t)$ denote steady-state measurement $x_t$ belongs to cluster $c_i$. $N_i$ denotes the count of observations $x_t \in c_i$.

$$\mathbf{1}_{c_i}(x) := \begin{cases} 1 \text{ if } x \in c_i \\ 0 \text{ if } x \notin c_i \end{cases} \quad (15)$$

$$N_i = \sum_{t=1:T} \mathbf{1}_{c_i}(x_t) \quad (16)$$

$$\sigma_i = sqrt\left(\frac{1}{N_i} \sum_{x \in c_i} (x - u_i)^2\right) \quad (17)$$

In summary, the HMM parameters $A$, $B$, $C$ and $\phi$ for event-driven NILM can be obtained using Eqs. (5-17). The emission vector $\phi$ of a refrigerator in LIFTED is shown in Table 1. The modified mean shift algorithm automatically clusters the refrigerator's measurement to four states. Physically, these four states are OFF, light on, motor on, and both motor and light on, respectively. That is, not only can the modified mean shift algorithm cluster the overall measurements into clusters, but each of the clustered states can be mapped to a real physical state of a given appliance. This means the proposed modified mean shift algorithm works well for type II appliances with multiple states, which is one contribution of this paper. Furthermore, the detected event segments all periods into different steady periods excluding transient periods. The partitioned steady periods are fed to the modified mean shift algorithm thus working well for type II appliance. It is one advantage of the proposed event-driven NILM.

**Table 1**
Emission vector of a refrigerator in LIFTED.

|   | State 1 | State 2 | State 3 | State 4 |
|---|---------|---------|---------|---------|
| μ | 0.52W | 42.03W | 121.62W | 156.60W |
| σ | 0.21W | 0.1W | 5.40W | 6.00W |

Based on the learned parameters $A$, $B$ and $C$, a Markov chain can be built. An example of the Markov chain for a refrigerator is shown in Fig. 4. Compared with the basic Markov chain, it enjoys two advantages. Firstly, it includes transient signatures DTS and DSP whose mean values are the second and third parameters in Fig. 4 and whose standard deviations are not listed. The transient signatures are unique to help label transitions of each appliance more accurately to identify appliances from the aggregated load. It should be noted that a power-decreasing transition is short, and it directly reaches another state without any transient spike power, so the DTS value for a power-decreasing transition is set to zero. Secondly, the states are not fully connected, so the transition matrix is sparse, and the computational cost is reduced. Furthermore, compared with a basic Markov chain, it does not include self-transition that transfers from one state to the state itself. While the number of transition paths in the basic Markov chain with 4 states is 16, that in the trained model for the refrigerator with 4 states is only 6, as shown in Fig. 4. The corresponding HMM inherits the sparsity from the Markov chain, so it is termed as sparse HMM.

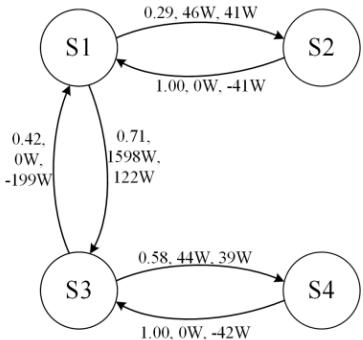

Fig. 4 Transition path and transition parameters of a refrigerator.

The sparsity of the Markov chain for an appliance can be defined as its sparsity level. For a basic Markov chain, the number of connected paths is $N_K = K^2$, where $K$ is the state number. Let $N_s$ be the number of connected paths in a trained Markov chain such as the one in Fig. 4. The sparsity level is $SL = \frac{N_s}{N_K}$. For example, the sparsity level of the Markov chain for the refrigerator in Fig. 4 is $SL = \frac{6}{16}$. The lower the sparsity level, the lower the computational complexity.

*2.4 Modeling a household as event-driven factorial HMM*

Assuming there are $N$ appliances in a household, the NILM problem can be formulated as a Factorial HMM (FHMM) by combining the HMMs of all individual appliances as shown in Fig. 5. Each appliance has its own chain to represent its working state $z_t^n$, denoting the state of appliance $n$ at time $t$. Given the composite load $x$, the task is to infer states of each appliance in different chains, i.e., $z_t^n$.

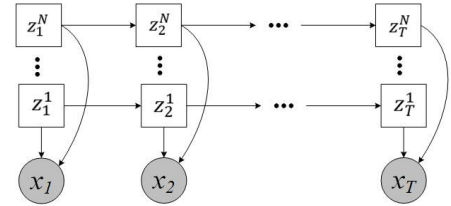

Fig. 5 Illustration of FHMM.

One standard approach of estimating the states of individual appliances is to maximize the likelihood of composite load $x$ and corresponding states $z$ as shown in Eq. (18). It enumerates the connected states of all the appliances within a household, which will have a high computational complexity when the number of time steps $T$ is large.

$$\hat{z}_{1:T} = \mathop{argmax}\limits_{z_{1:T}} p(x_{1:T}, z_{1:T}) \qquad (18)$$

The probability $p(x_{1:T}, z_{1:T})$ can be further formulated as Eq. (19) based on $\theta = \{A, B, C, \phi\}$, the learned HMM parameters for individual appliances.

$$\begin{aligned} p(x_{1:T}, z_{1:T}) &= p(x, z|\theta) \\ &= \prod_{t=2}^{T} p(z_t|z_{t-1}, A) \times \prod_{t=2}^{T} p(x_t|z_t, \phi) \\ &\propto \prod_{m=1}^{M} p(z_{m+1}|z_m, e_m, A, B, C) \times \prod_{m=1}^{M} p(\varphi_{m+1}|z_{m+1}, \phi) \end{aligned} \qquad (19)$$

Since the state changes only when an event occurs, it is not necessary to perform state inference from $t$-1 to $t$ in the steady period. Accordingly, $p(z_t|z_{t-1}, A)$ can be converted to $p(z_{m+1}|z_m, e_m, A, B, C)$, where $m$ is the index of events, $z_m$ and $z_{m+1}$ refer to the states of the $m^{th}$ and $(m+1)^{th}$ steady periods, respectively, before and after the $m^{th}$ event, and $e_m$ represents the extracted transient signatures DTS and DSP from the $m^{th}$ transition process. $\varphi_{m+1}$ is the extracted steady-state signature SSP of the $(m+1)^{th}$ steady period.

The above reformulation further naturally improves NILM model of multiple appliances in a household. The basic FHMM aims to infer states of individual appliances in a point-to-point way. If the sampling rate is high or if the appliance number is big, it will be very difficult, if not impossible, to perform inference due to the high computational complexity. However, the event-driven solution just identifies state-changing, appliances when an event occurs thus the basic FHMM can be transformed to the event-driven FHMM (eFHMM) as shown in Fig. 6. It achieves online identification no matter how large the sampling rate or appliance number is.

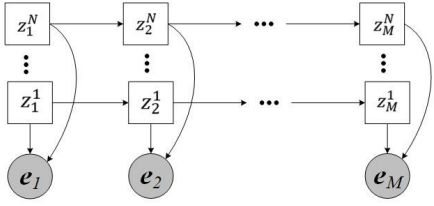

Fig. 6 Illustration of eFHMM.

### 3. Two-stage Solution to Event-driven Factorial HMM

Reformulation of the basic FHMM into the event-driven FHMM significantly decreases the computational complexity by transferring point-to-point inference to the event-driven inference. The performance of load disaggregation can be further improved by a two-stage NILM solution, termed as eFHMM-TS, which uses the extracted transient and steady-state signatures specifically designed for the proposed eFHMM. Computational complexity of the two-stage NILM solution for eFHMM will be described in detail.

*3.1 Two-stage NILM solution to eFHMM*

The reformulation of FHMM can decrease computational complexity significantly; however, it still performs factorials as the product of probabilities in Eq. (19). For ease of computation, it can be further converted to the summation of logarithms of probabilities at different times for different appliances as shown in Eq. (20).

$$\mathcal{L}(x, z|\theta) = \sum_{m=1}^{M} \log p(z_{m+1}|z_m, e_m, A, B, C) + \sum_{m=1}^{M} \log p(\varphi_{m+1}|z_{m+1}, \phi) \quad (20)$$

One contribution of the proposed eFHMM-TS method is to achieve real-time appliance identification. Using transient signatures, the most likely state-changing appliance can be inferred by calculating the probability of each state transition path of individual appliance. Thus, Eq. (20) is further developed into Eq. (21) by considering individual appliances and finding the maximum probability of each transition path introduced in appliance Markov chains. It should be noted that in HMM-based method, the current state $z_m^n$ is known, so it can start from this state to find next states along its connected paths.

$$\max_{n=1:N} \mathcal{L}(x, z^{1:N}|\theta^{1:N})$$
$$= \sum_{m=1}^{M} \max_{n=1:N} \log p(z_{m+1}^n|z_m^n, e_m, A^n, B^n, C^n)$$
$$+ \sum_{m=1}^{M} \max_{n=1:N} \log p(\varphi_{m+1}|z_{m+1}^{(1:N)}, \phi^{(1:N)}) \quad (21)$$

Since each appliance is independent of other appliances and each event is independent of other events as well, the local maximum probability of each appliance for each event would guarantee the global maximum. The probability of appliance $n$ in the second row of (21) can be further expanded to Eq. (22).

$$p(z_{m+1}^n|z_m^n, e_m, A^n, B^n, C^n)$$
$$= \frac{p(z_{m+1}^n, z_m^n, e_m, A^n, B^n, C^n)}{p(z_m^n, e_m, A^n, B^n, C^n)}$$
$$= \frac{p(e_m|z_{m+1}^n, z_m^n, A^n, B^n, C^n)}{p(e_m|A^n, B^n, C^n, z_{m-1}^n)}$$
$$\times \frac{p(z_{m+1}^n, z_m^n, A^n, B^n, C^n)}{p(A^n, B^n, C^n, z_m^n)}$$
$$= \frac{p(e_m|z_{m+1}^n, z_m^n, A^n, B^n, C^n)}{p(e_m|A^n, B^n, C^n, z_{m-1}^n)}$$
$$\times p(z_{m+1}^n|z_m^n, A^n, B^n, C^n)$$
$$\propto p(e_m|z_{m+1}^n, z_m^n, B^n, C^n)$$
$$\times p(z_{m+1}^n|z_m^n, A^n) \quad (22)$$

Since the first row in (21) is a conditional probability, it can be further developed using Bayesian inference based on the available evidence, i.e., transient signatures, $e_m$. It finally becomes the last two rows in (22) after derivation. The last row is the transition probability of appliance $n$. The second last row is the probability of the transient signatures $e_t$ belonging to appliance $n$. It starts from the current state of appliance $n$ and goes to its connected states and calculates the probability using learned Gaussian distribution $B^n, C^n$. Take Fig. 4 as an example and assume that the current state is S3. The connected states of S3 are S1 and S4. If the mean value of the extracted transient signature DSP is positive, it indicates a power-increasing change and will only calculate the probability from S3 to S4 rather than S3 to S1. In addition, it will never calculate the probability between S3 and S2, because these two states are not directly connected. After enumerating all the appliances, the calculated probability will be sorted and the transition with the maximum probability will be chosen as the true state. The associated appliance will be the state-changing appliance.

The last row in Eq. (21) is the probability of steady-state signature $\varphi$ extracted from the $(m+1)^{th}$ steady period given the states $z_{m+1}^{(1:N)}$ of individual appliances and their associated learned parameters $\phi^{(1:N)}$. It has not been clear how to use it so far. The only thing that can be done is to express it using the signatures of individual appliances. The composite load can be expressed as the summation of ground truth of individual appliances. Since states of all the appliances are independent and identically distributed, the equation can be converted from the second row to the third row in Eq. (23), where $n_s$ is state $s$ of appliance $n$.

$$\max \log p(\varphi_{m+1}|z_{m+1}^{(1:N)}, \phi^{(1:N)}) \quad (23)$$

$$= \max_{n=1:N} \left( -\sum_{n=1}^{N} \log \sqrt{2\pi}\sigma_{n_s} - \frac{1}{2}\sum_{n=1}^{N} \left( \frac{\varphi_{m+1}^n - \mu_{n_s}}{\sigma_{n_s}} \right)^2 \right)$$

$$\propto \min_{n=1:N} \sum_{n=1}^{N} \left( \frac{\varphi_{m+1}^n - \mu_{n_s}}{\sigma_{n_s}} \right)^2$$

$$\propto \min_{n=1:N} \sum_{n=1}^{N} |\varphi_{m+1}^n - \mu_{n_s}|$$

According to the quantile function of Gaussian distribution, random variable $\varphi_{m+1}^n$ will lie inside the interval $\mu_{n_s} \pm 2.576\sigma_{n_s}$ with probability 99% as shown in Eq. (24). The summation of the absolute differences between ground truth and mean of SSP for all the appliances is shown in Eq. (25). The rightmost part $\varphi_{m+1} - \sum_{n=1}^{N} \mu_{n_s}$ in (27) should be less than $\sum_{n=1}^{N} 2.576\sigma_{n_s}$.

$$|\varphi_{m+1}^n - \mu_{n_s}| \leq 2.576\sigma_{n_s} \tag{24}$$

$$\sum_{n=1}^{N} 2.576\sigma_{n_s} \geq \sum_{n=1}^{N} |\varphi_{m+1}^n - \mu_{n_s}| \geq \sum_{n=1}^{N} (\varphi_{m+1}^n - \mu_{n_s})$$

$$= \varphi_{m+1} - \sum_{n=1}^{N} \mu_{n_s} \tag{25}$$

Thus, the derivation in Eq. (25) shows that the third row in Eq. (21) just applies the steady-state signature SSP. It can be used to judge whether the inferred states in the first stage is correct or not by comparing the difference between the composite load and the summation of mean value of current state. These two stages will work iteratively to find the most likely appliance. If the inferred states in first stage violate Eq. (25), the appliance with the second highest probability returned from Eq. (22) will be regarded as state-changing appliance and further verified in Eq. (25). It will iteratively find the solution in this way to meet requirement of Eq. (25). Therefore, it is a two-stage NILM solution for the eFHMM, where the first stage is to infer the changing state and the second state is to judge whether the inferred states is correct or not.

*3.2 Computational complexity*

The computational complexity for Eq. (19) could be about $O(TK^{2N})$ in theory, where $K$ is the average number of states of each appliance, $N$ is the number of appliances, and $T$ is the total length or the number of electrical measurements. This problem is intractable due to its high computational complexity. Suppose there are 10 appliances and the average number of states of each appliance is 3. Even for two timesteps, the computational cost is $3^{20}$, which is too large to implement in real-time load disaggregation. Using logarithm in Eq. (20), the complexity decreases to $O(TNK^2)$ that can use exact inference to identify appliances, but it is still hard to perform online NILM.

The implementation of event-driven NILM significantly decreases the inference times from the number of observations T to the number of events $M$, which is limited even for a long time, i.e., $M \ll T$. The computational complexity decreases to $O(MNK^2)$. The number of events $M$ is immune to the sampling rate, so online disaggregation is possible even for high-resolution data.

As individual appliances are modeled as sparse HMM in which not all the states are interconnected, the computational complexity can further decrease depending on the sparsity level. In Fig. 4, the sparsity level of the refrigerator's HMM is $\frac{3}{8}$, so its computational complexity roughly becomes $O(\frac{3}{8}MNK^2)$. In eFHMM, given the current state of each appliance, it starts from the current state to determine the next state. Based on it, the computational complexity decreases to $O(\frac{3}{8}MNK)$, i.e., $O(MNK)$. Therefore, computational time is linear to the number of events, the number of appliances, and the average number of appliance states.

**4. Integrating eFHMM-TS into Edge-cloud Framework**

Typical NILM methods require high computational capacity to deal with complex processes. It is hard to disaggregate the composite load on the end side that has limited computational capacity, or it is a huge resource waste even if it is done on the end side that has ample computational capacity. The centralized model, i.e., using cloud service, faces three significant challenges: 1) Congestion of big data: in centralized model, massive data needs to be transferred from end devices to remote cloud servers, so it is inevitable to consider data congestion due to the unstable communication system; 2) Latency between end devices and cloud server: high latency can be a big challenge when end devices are relatively far away from cloud server. 3) Privacy: it exposes the raw data to public network giving outsiders access to data and causing unauthorized intrusion of data.

The above challenges can be addressed with an edge-cloud framework, in which multiple edge servers bridge the gap between IoT devices and cloud infrastructure, and the cloud server takes over the task of load disaggregation. First, by providing computing capacity, storage space and available communication bandwidth to assist the resource-poor IoT devices, the IoT devices can offload their burdens to the edge servers. Second, since the edge servers are closer to the IoT devices, the latency can be well controlled compared to the centralized model. Third, only a small part of information is exposed in the public network, there is less risk on unauthorized disclosure of data.

The event-driven two-stage NILM solution is naturally suited for the edge-cloud framework. Event detection and signature extraction are tasks that require limited computing resources so they can be completed in edge servers and the extracted signatures are transmitted to the cloud server to complete more sophisticated and computation-intensive load disaggregation. Based on the appliance-level details, utility company could provide more novel and better services such as load forecasting and shedding.

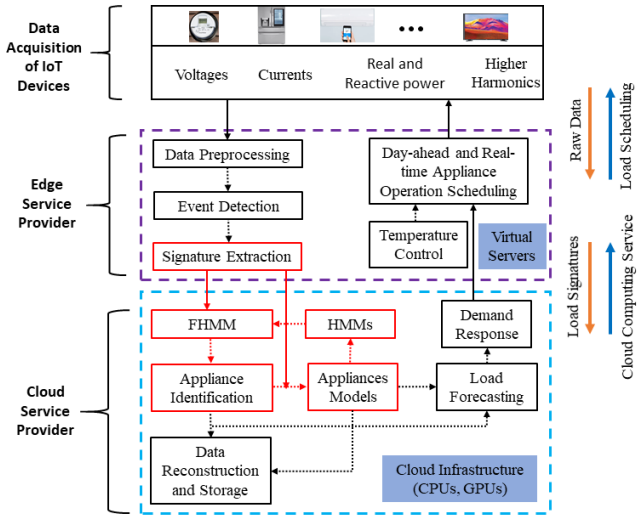

Fig. 7 The proposed edge-cloud framework for event-driven NILM.

Fig. 7 outlines the proposed edge-cloud framework from data acquisition to load scheduling in which the red part is covered in this paper. The proposed eFHMM-TS is the most important part from IoT devices to cloud infrastructure. Each layer of Fig. 7 is responsible for different tasks.

(1) *Data acquisition of IoT devices*. With the development of IoT technology, some commonly used home appliances such as TV, air conditioner and heater could upload their power usage data to cloud server and be remotely controlled. These devices could provide ground truth of individual appliances which will decrease the difficulty of load disaggregation. The Advanced Smart Meters convey the composite load information of all the home appliances. Such information would be sent to the edge server even they are likely in different resolution.

(2) *Edge service provider*. Edge servers are some virtual servers that can manage data from multiple buildings in one computer at the same time. Due to the limited computational capacity assigned to one building, it only executes simple task. It completes the data preprocessing at first including dealing with missing values and outliers. Event detection and load signature extraction are also completed in edge servers. Accordingly, the size of data can be reduced significantly when signatures instead of raw data are transmitted to cloud server.

(3) *Cloud service provider*. Cloud infrastructure contains lots of CPUs and GPUs that can perform complex computation. Load disaggregation of NILM will be executed on cloud side. The appliance model will be learned, and appliance model library will be assembled in the cloud. For the load disaggregation of a given house, the corresponding appliance HMMs will be retrieved from the model library. Once done with the appliance identification, some advanced applications such as data reconstruction, load forecasting and demand response will be performed. The outcome of these applications will be provided to help guide users' green behaviors.

## 5. Case study

Testing on high-resolution datasets with appliance-level ground truth can verify how eFHMM-TS takes advantage of event detection and transient information to improve performance. In this section, experiments will be conducted on three datasets with different resolutions including 50 Hz LIFTED, 10 Hz LIFTED, and 5 Hz synD to verify the event-driven two-stage solution method (eFHMM-TS) and compare the eFHMM-TS method with the SIQCP method and different methods in NILMTK. All the solution algorithms are coded in Python 3.6 running on a desktop with 4.2GHz intel i7-7700K processor and 16GB memory.

### 5.1 Testing datasets and evaluation metrics

LIFTED is a new one-week NILM dataset down sampled at 50 Hz [32]. It contains appliance-level details including voltage, current, active power and reactive power for electrical appliances. In this paper, we test the proposed eFHMM-TS method and comparing methods on 10 appliances, namely, kettle, steamer, toaster, hotpot, vacuum, hair dryer, refrigerator, mixer, blender, washing machine. The last six of them are motor-driven appliances with obvious transient signatures. In practice, the difficulty of load disaggregation depends on the similarity of appliances and how appliances are mixed at the same time. So, we randomly mix 10 appliances for a period of about 10 hours while continuously switching appliance states to ensure the complexity of the test cases. The aggregated power is shown in Fig. 8.

To further compare eFHMM-TS with other reported methods, tests on the synthetic data of LIFTED down sampled to 10 Hz and the 5 Hz synD dataset released in April 2020 are also conducted. SynD is a publicly available dataset [33]. It provides ground truth for appliances that can be used to test NILM methods. In this paper, fridge, kettle, toaster, hair dryer, coffee maker, fan, and electric space heater that consume more energy are selected. Compared with the previous two datasets (50Hz LIFTED and 10Hz LIFTED), the transient processes in synD are not stable due to its low sampling rate that brings challenges for the proposed method. Take fridge as an example. As shown in Fig. 9, its transient spike power is about from 630w to 930w. Another difference is that most appliances in synD only turn ON and OFF limited times in one day that leads to a low RMSE even though appliances are not identified at all. The data of first 30 days is used for training and the following 20 days is used for testing.

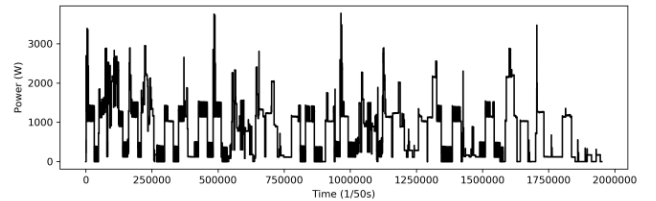

Fig. 8 Aggregated power of synthetic test case in LIFTED.

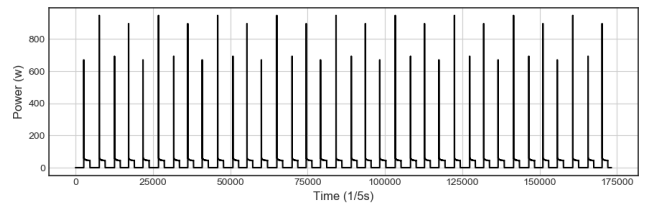

Fig. 9 Cycles of fridge in synD dataset.

To evaluate the performance of eFHMM-TS and other methods, metrics including accuracy, precision, recall and f1-score are used as shown in Eqs. (25-28), where TP, TN, FP, and FN represent true positive, true negative, false positive, and false negative, respectively. These four metrics reflect appliance state at ON or OFF. When an appliance is on, it will be counted as 1 no matter which state it is. Otherwise, it will be counted as 0. Besides, root mean square error (RMSE) is used to evaluate the gap between estimated and true power consumption as in Eq. (29), where $p_i$ and $\widehat{p}_i$ are the $i^{th}$ true and estimated power consumption, respectively, and $l$ is the length of power measurements.

$$Accuracy = \frac{TP+TN}{TP+TN+FP+FN} \quad (25)$$
$$Precision = \frac{TP}{TP+FP} \quad (26)$$
$$Recall = \frac{TP}{TP+FN} \quad (27)$$
$$f_1 = \frac{2 \times Precision \times Recall}{Precision+Recall} \quad (28)$$
$$RMSE = \sqrt{\frac{\sum_i (p_i - \widehat{p}_i)^2}{l}} \quad (29)$$

*5.2 Performance evaluation of eFHMM-TS*

The proposed eFHMM-TS method applies transient signatures to identify state-changing appliances and verify if the inferred state in first stage is correct or not in the second stage using steady-state signatures. This section is to validate the effectiveness of the first and second stages by removing transient and steady-state signatures in the two-stage NILM solution.

Three algorithms are used to validate the effectiveness of the combination of transient and steady-state signatures in the full two-stage algorithm (FULL). Full algorithm without transient signatures (FWT) in the first stage and full algorithm without steady-state signatures (FWS) in the second stage are used for comparison. Fig. 10 shows the accuracy, precision, recall and f1-score of the three algorithms.

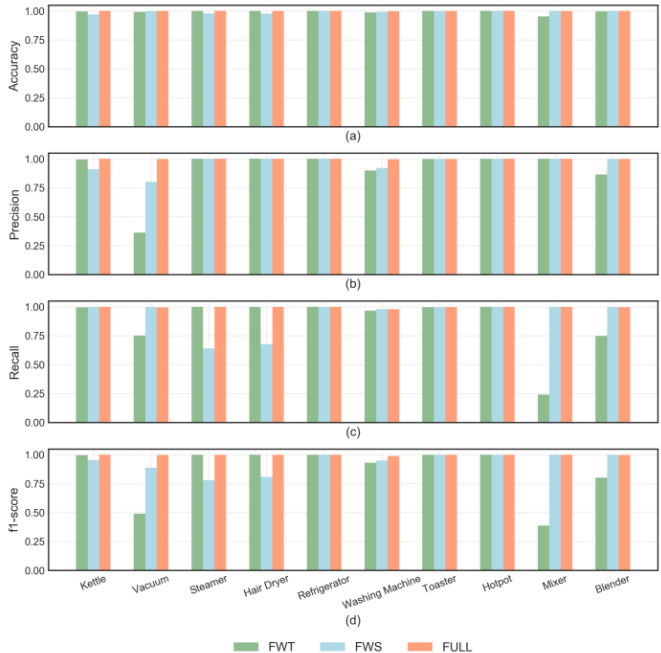

Fig. 10 Comparison of FWT, FWS and FULL algorithms, (a) Accuracy, (b) Precision, (c) Recall, (d) f1-score.

**Table 2**
Sample mean and standard deviation of DTS and DSP.

|  | DTS | | DSP | |
|---|---|---|---|---|
| Appliance | μ | σ | μ | σ |
| Kettle (s1->s2) | 1053.8 | 4.8 | 1026.3 | 5.2 |
| Vacuum (s1->s2) | 2387.3 | 58.3 | 1001.8 | 17.3 |
| Hair Dryer (s2->s5) | 1246.4 | 27.1 | 1194.1 | 17.4 |

**Table 3**
RMSE comparison of FWT, FWS and FULL algorithms.

|  | FWT | FWS | FULL |
|---|---|---|---|
| Kettle | 55.4 | 178.8 | 8.9 |
| Vacuum | 95.5 | 38.6 | 8.2 |
| Steamer | 3.26 | 113.4 | 3.3 |
| Hair Dryer | 97.4 | 180.9 | 6.9 |
| Refrigerator | 29.5 | 13.4 | 11.9 |
| Washing Machine | 27.5 | 13.2 | 10.7 |
| Toaster | 4.2 | 4.2 | 4.2 |
| Hotpot | 6.2 | 6.2 | 6.2 |
| Mixer | 8.6 | 0.96 | 0.9 |
| Blender | 13.9 | 2.3 | 2.3 |

It should be noted that two signatures, i.e., DTS and DSP, are used in the first stage. Even though DSP is extracted from transient process, it expresses steady-state information that can be easily extracted even from low-resolution data, and it is used to infer changing state in the first stage for FWT. Nevertheless, FWT still struggles to correctly identify vacuum and mixer mainly for two reasons: 1) load signature similarity among different appliances, and 2) high fluctuation of composite load. It is hard to accurately extract signatures in composite load due to its high fluctuation. Even though all the signatures are modeled as Gaussian distribution, the value of one signature extracted from composite load may exceed three standard deviations from the learned mean. In such case, load signature similarity will greatly affect identification accuracy. Table 2 lists samples of DTS and DSP of kettle, vacuum, and hair dryer.

While it is hard to accurately distinguish kettle, vacuum and hair dryer by solely using DSP, the DTS could help further separate these appliances due to its significant difference. For example, the distribution of mean of DSP between kettle and vacuum overlaps each other, and the extracted DSP from state 2 to state 5 of hair dryer may be close to these two appliances in measured power. RMSE of these three appliances is 55.4w, 95.5w and 97.4w using FWT as shown in Table 3, which means these three appliances are falsely identified. Performance metrics of hair dryer using FWT are bigger than 0.9 in Fig. 10. That is because no matter which state it is at, it is counted as ON. That may be one drawback of the performance metrics when used for multi-state appliances. For small appliances such as mixer which have small RMSEs, metrics including accuracy, precision, recall and f1-score can better reflect the performance of different methods. It can be observed that, without transient in FWT, it is hard to identify the working states of mixer, which proves that transient signatures play an important role in accurately identifying appliances. The second stage is to confirm whether the estimated states are correct or not by comparing the difference of total estimated load using SSP and measured composite load. For HR-based appliances, their turn-on and turn-off events do not have transient information, it easily leads to misidentification. For motor-driven appliances, their turn-off events do not have transient information either. That is why kettle, vacuum, hair dryer and steamer have poor RMSE in FWS. Thus, the second stage is also important to accurately identify appliances and methods such as SIQCP in [18] that highly depend on steady-state signatures should have similar problems.

*5.3 Performance comparison with SIQCP*

This section presents the performance comparison between the proposed eFHMM-TS with SIQCP on LIFTED dataset that aims to illustrate the reasons of worse performance of SIQCP. To the best of our knowledge, SIQCP is one of the best NILM methods of its kind. The ratio of training and testing data is 5:5. As mentioned before, SIQCP highly depends on SSP, and it proposes an iterative k-means method to determine state number of individual appliances and associated SSP.

The performances for cases with different appliance numbers are shown in Fig. 11. It shows that the performance of SIQCP decreases with increasing number of appliances because it only uses steady-state signatures, and some appliances have similar steady-state signatures. More appliances mean higher probability of signature similarity. Thus, with the increase of the appliance number, the performance will decrease accordingly. For event-driven method, it takes advantage of transient signatures which are different for different appliances that can help separate them.

Among all the evaluation metrics, accuracy is comparatively high because most of home appliances are OFF at most time, and even if one appliance is misidentified, the misidentification will end when the associated appliance turns off or switches mode. Recall, precision and $f_1$-score of SIQCP decrease greatly when the number of appliances increases. Apart from similar signatures, the determination of the number of states is another factor that affects the performance. Table 4 lists the state number and corresponding mean $\mu$ and standard deviation $\sigma$ of SIQCP and eFHMM-TS.

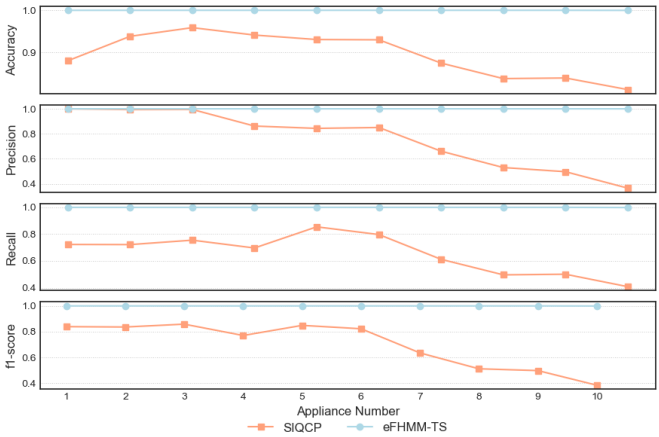

Fig. 11 Performance comparison between SIQCP and eFHMM-TS with increasing appliance number.

**Table 4**
Mean and standard deviation of determined states in SIQCP and eFHMM-TS.

| Appliance | State No. | SIQCP $\mu$ | SIQCP $\sigma$ | eFHMM-TS $\mu$ | eFHMM-TS $\sigma$ |
|---|---|---|---|---|---|
| Kettle | 1 | 0.4 | 2.6 | 0.4 | 0.6 |
| | 2 | 1026.8 | 6.4 | 1027.1 | 5.1 |
| Vacuum | 1 | 0.01 | 1.4 | 0.16 | 0.01 |
| | 2 | 1002.6 | 26.9 | 1001.8 | 22.1 |
| | 3 | 1315.4 | 112.7 | | |
| | 4 | 1805.1 | 140.2 | | |
| | 5 | 2276.5 | 108.5 | | |
| Hair Dryer | 1 | 0.01 | 0.2 | 0.3 | 0.1 |
| | 2 | 99.6 | 23.4 | 80.7 | 1.2 |
| | 3 | 686.9 | 8.1 | 127.9 | 1.4 |
| | 4 | 1015.9 | 103.7 | 687.7 | 2.8 |
| | 5 | 1370.6 | 4.7 | 1371.1 | 3.6 |
| Refrigerator | 1 | 0.01 | 0.2 | 0.5 | 0.2 |
| | 2 | 119.2 | 5.7 | 172.6 | 2.4 |
| | 3 | 1105.6 | 130.2 | 255.5 | 9.8 |
| | 4 | 1540.7 | 78.3 | 317.2 | 9.6 |
| Blender | 1 | 0.1 | 0.3 | 0.2 | 0.1 |
| | 2 | 226.2 | 4.8 | 226.3 | 3.9 |
| | 3 | 294.1 | 25.5 | | |
| | 4 | 404.1 | 32.1 | | |
| | 5 | 523.6 | 30.6 | | |

As an example, vacuum has two actual states: ON and OFF. However, based on the iterative k-means method in [18], five states are identified for the same vacuum. This clustering method will cause the following problems:

- More states increase the probability of load signature similarity thus leading to appliance misidentification.
- More states lead to more paths in Markov chains as shown in Fig. 4, which makes each transition probability smaller (summation of all paths beginning from the same state is 1). It may further decrease the inference probability even for the true state-changing appliance.
- The identified states are statistical results due to the inappropriate clustering method. Physically, it does not have much practical significance, and cannot represent appliances in an accurate fashion. Take refrigerator as an example, even though it has the same state number as our method, it clusters the transient spike that is short but not a true state as two states, i.e., 1015.6W and 1540.7W in Table 4.

Due to the wrongly learned parameters and associated load signature similarity, SIQCP cannot achieve a good performance on individual appliances when testing 10 appliances, as shown in Fig. 12. In comparison, eFHMM-TS accurately detects each appliance. Precision refers to true estimated ratio of ON state. The precisions using SIQCP on vacuum, mixer, blender and hair dryer are as low as 0.41, 0.77, 0.53 and 0.16, respectively, whereas the precisions using eFHMM-TS are 0.99, 0.99, 0.99 and 0.99, respectively, which means eFHMM-TS identifies these appliances almost perfectly. As an example, the SSPs of vacuum and kettle overlap each other as mentioned before, which is hard to distinguish using nonevent-based algorithms. As shown in Fig. 12, the recall and f1-score of SIQCP are much lower than those of eFHMM-TS.

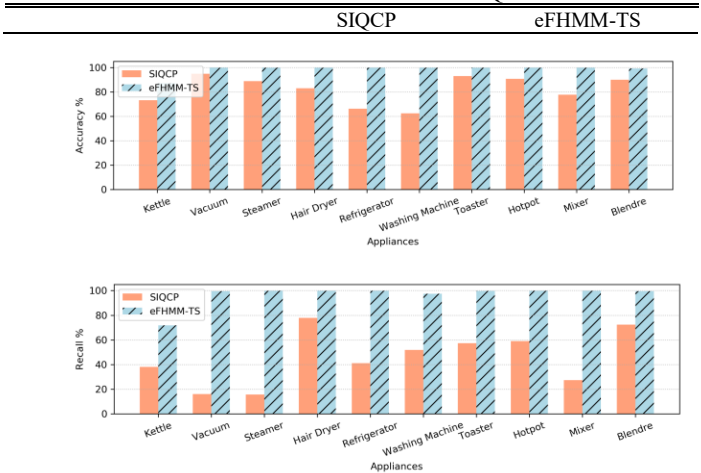
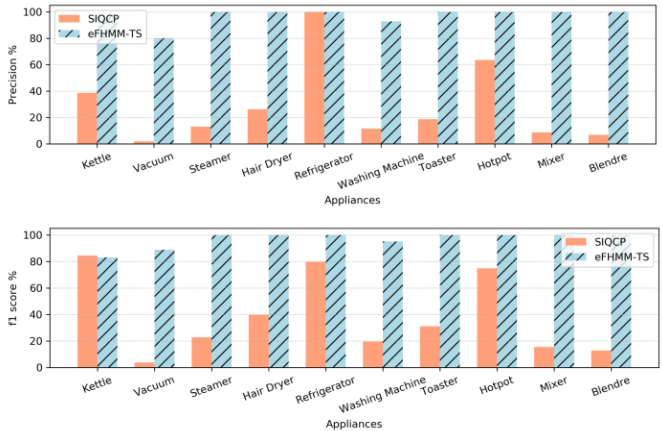

Fig. 12 Performance comparison on individual appliances between SIQCP and eFHMM-TS.

*5.4 Performance comparison with methods in NILMTK*

NILMTK v0.4 was released in August 2019. It is used to compare the proposed two-stage NILM solution, eFHMM-TS, on LIFTED in this section. NILMTK includes combinatorial optimization (CO) [3], FHMM_EXACT [19, 34], Hart_85 [35], and Mean [36] methods. HART_85 is one of the most used baseline NILM methods with edge detection. CO is similar to subset sum problem that aims to minimize the difference between the household aggregate reading and the sum of power usages of different appliances. Super HMM of FHMM-EXACT is to represent the states of each appliance to identify appliances from aggregated power. Mean method predicts all appliances to be ON and returns the mean power value for all appliances. The comparison results in terms of accuracy, precision, recall, f1-score are shown in Fig. 13. RMSE comparison among these methods is listed in Table 5.

In terms of accuracy, precision, recall and f1-score, the proposed eFHMM-TS outperforms the other three methods, while the Mean method has very poor performance, so it is not listed here. A threshold of 15w is applied to Hart_85 in this paper for edge detection. For transition process, the difference between two adjacent datapoints is usually larger than 15w so Hart_85 misidentifies lots of events. The same cases are from high fluctuation that can easily generate more events and lead to false appliance identification. CO determines the state number of individual appliances before clustering, so it leads to false state number and associated power. In fact, it is not true to force all the appliances having the same state number. For example, the kettle has two states: ON and OFF. If the state number is forced to be three, the means of the three states using CO are 0w, 320w, 1026w in which the state with a mean of 320w is not correct.

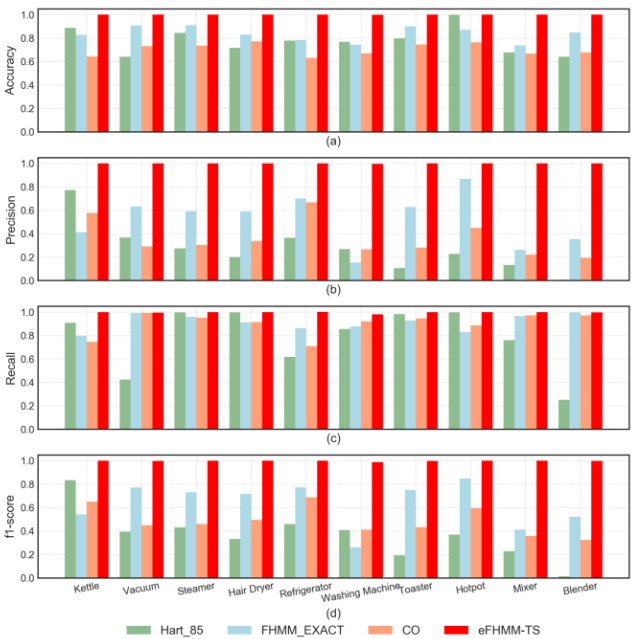

Fig. 13 Performance comparison among different methods on 50 Hz LIFTED. (a) Accuracy, (b) Precision, (c) Recall, (d) f1-score.

**Table 5**
RMSE comparison among Hart_85, CO, FHMM_EXACT and eFHMM-TS on 50 Hz LIFTED.

|  | Hart_85 | CO | FHMM_EXACT | eFHMM-TS |
|---|---|---|---|---|
| Kettle | 345.8 | 383.2 | 306.3 | 8.9 |
| Vacuum | 105.1 | 163.3 | 82.2 | 8.2 |
| Steamer | 301.7 | 180.4 | 144.1 | 3.3 |
| Hair Dryer | 199.2 | 154.9 | 155.2 | 6.9 |
| Refrigerator | 55.7 | 110.2 | 41.7 | 11.9 |
| Washing Machine | 122.3 | 101.3 | 80.9 | 10.7 |
| Toaster | 325.7 | 149.6 | 126.9 | 4.2 |
| Hotpot | 16.1 | 244.2 | 206.4 | 6.2 |
| Mixer | 13.9 | 18.6 | 15.4 | 0.9 |
| Blender | 45.1 | 34.2 | 62.3 | 2.3 |

Besides, combination of appliance states in CO may have a natural drawback, which in itself may be a multi-solution problem. The combination of false estimated states may lead to the same composite load. These problems exist in FHMM_EXACT as well which leads to poor performance. However, since it applies sparse HMM and transition probability between different states that are ignored in Hart_85 and CO, its performance is much better. In terms of f1-score, FHMM_EXACT outperforms Hart_85 and CO on 8 out of 10 appliances in LIFTED. The average f1-scores of Hart_85, CO, and FHMM_EXACT are 0.36, 0.48, and 0.63, respectively. However, the f1-score of the proposed eFHMM-TS is as high as 0.98, which is far better than any method in NILMTK.

The precisions of Hart_85, CO, and FHMM_EXACT are all low. Take FHMM_EXACT as an example, the precisions of individual appliances are 0.41, 0.63, 0.59, 0.58, 0.70, 0.15, 0.62, 0.86, 0.26, and 0.35. It means that there are many false positive estimations because of signature similarity and false estimated state of individual appliances. These are also the true reasons why these methods have high RMSE, especially for appliances with high power consumption such as kettle and hair dryer.

*5.5 Performance evaluation on 10 Hz dataset*

To further compare eFHMM-TS with different methods in NILMTK and SIQCP, the 50Hz LIFTED dataset is down sampled to 10Hz for testing in this section. It is more difficult to identify each appliance on lower-resolution dataset, especially in this case due to the near-simultaneous events caused by washing machine's frequently turning ON and OFF in wash mode. The high fluctuation easily masks small appliances and increases segment number thus leading to more inference times that may decrease identification accuracy. The performance of different methods is shown in Fig. 14.

As shown in Fig. 14, f1-score of Hart_85, CO, Mean FHMM_EXACT, and SIQCP on washing machine is as low as 0.20, 0.39, 0.09, 0.10, and 0.34, respectively. This is because washing machine is a motor-driven appliance that has transient spike power, which is easy to be clustered as a state in the training step in those methods; however, the transient spike is too short to identify in load disaggregation step. NILMTK configures two parameters ('chunck size' and 'sample rate') to disaggregate load in preset segments and accelerate computation speed. Because of these settings, two different states that may occur in one segment are forced to be identified as one state.

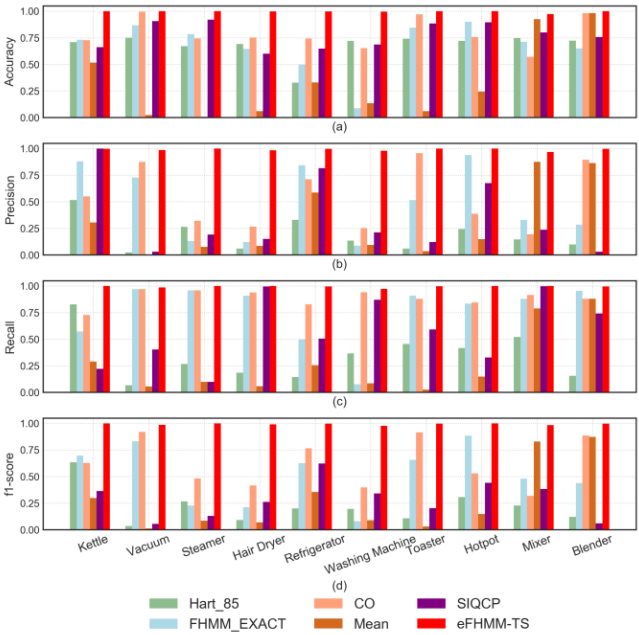

Fig. 14 Performance comparison among different methods on 10 Hz LIFTED. (a) Accuracy, (b) Precision, (c) Recall, (d) f1-score.

It shows that those methods are not suitable for the appliances with obvious transient spike even on low-resolution datasets. Since different methods adopt different load disaggregation algorithms with different characteristics, they have different performance on different kinds of appliances. For example, Hart_85 struggles to detect vacuum, hair dryer and toaster; CO fails to identify vacuum, steamer and hair dryer; SIQCP misidentifies vacuum, steamer and blender, etc. In comparison, the proposed eFHMM-TS method achieves good performance on each appliance.

RMSE of different methods on individual appliances is shown in Table 6. Since every appliance works only for several cycles in about 5.4 hours in the synthetic dataset, it will lead to high RMSE if the appliances are not identified accurately due to the short working time $l$ in Eq. (31).

**Table 6**
RMSE comparison among Hart_85, FHMM_EXACT, CO, Mean, SIQCP and eFHMM-TS on 10 Hz LIFTED.

|  | Hart_85 | CO | FHMM_EXACT | Mean | SIQCP | eFHMM-TS |
|---|---|---|---|---|---|---|
| Kettle | 235.4 | 353.5 | 349.1 | 457.4 | 377.5 | 26.7 |
| Vacuum | 87.2 | 6.3 | 79.8 | 38.4 | 215.7 | 12.6 |
| Steamer | 208.5 | 269.5 | 249.9 | 191.3 | 136.9 | 4.9 |
| Hair Dryer | 218.7 | 204.7 | 273.4 | 200.8 | 225.6 | 20.1 |
| Refrigerator | 107.8 | 52.9 | 68.9 | 57.9 | 221.8 | 12.7 |
| Washing Machine | 85.8 | 62.4 | 88.1 | 95.9 | 110.3 | 15.4 |
| Toaster | 117.1 | 31.4 | 110.7 | 94.9 | 142.1 | 5.6 |
| Hotpot | 389.8 | 279.1 | 164.2 | 379.2 | 277.6 | 15.3 |
| Mixer | 81.3 | 16.9 | 14.7 | 9.6 | 16.5 | 6.3 |
| Blender | 85.5 | 6.4 | 57.2 | 19.1 | 61.3 | 2.6 |

*5.6 Performance evaluation on synD dataset*

In this section, we use the 5Hz synD dataset to further evaluate the performance of eFHMM-TS, different methods in NILMTK, and SIQCP. As shown in Fig. 15, FHMM_EXACT and CO achieve comparatively good performance because they force to cluster each appliance as two states. It is just suitable for the chosen appliances. Meanwhile, different appliances work at different time slots, and it is easy to infer working appliances using combination optimization. When there are not near-simultaneous events, it is of high performance using FHMM-based method. However, SIQCP determines state number of each appliance according to two preset parameters '*maximum state number*' and '*standard deviation of each state*'. It makes state clustering towards the preset maximum state number when there are transient spike power and high fluctuation, which leads to misidentification in load disaggregation. Even though it struggles to identify appliances on high-resolution datasets, it works well for low-resolution dataset.

On the low-resolution dataset, the proposed eFHMM-TS does not have overwhelming superiority over other methods, as is the case in the high-resolution dataset. It is partly because transient signatures of some appliances are not stable when they appear, and standard deviations of some transient signatures are fairly big, which may lead to misidentification in the first stage of eFHMM-TS. Despite that, eFHMM-TS still has one advantage other methods, which is to rectify the falsely identified appliances in the second stage until the estimated power consumption meets the requirement in Eq. (27). In terms of accuracy, precision, recall and f1-score, the average values of eFHMM-TS are 0.99, 0.94, 0.94, 0.94, and those of FHMM_EXACT is 0.96, 0.92, 0.97 and 0.95. That is, the recall and f1-score of FHMM_EXACT are higher (0.03 and 0.01) than eFHMM-TS. As shown in Fig. 15, eFHMM-TS cannot fully identify fridge due to its unstable transient signatures as shown in Fig. 9, which is the main reason for the lower overall performance.

Mean method does not consider either state transition, as in SIQCP and FHMM_EXACT, or the combination of appliances, as in CO, so it is hard to achieve a good disaggregation performance. Its performance is worse than that of CO or FHMM_EXACT. As it assumes that appliances are "always on", it may only suit to identify one appliance from multiple ones.

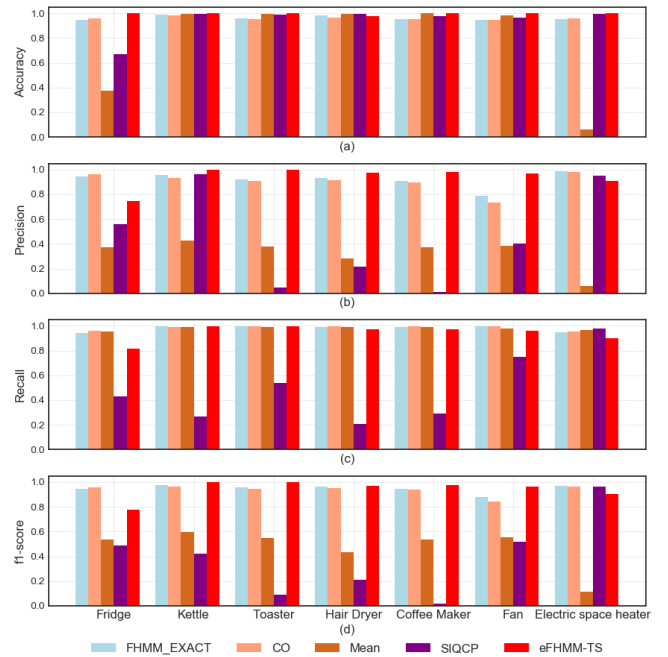

Fig. 15 Performance comparison among different methods on synD dataset. (a) Accuracy, (b) Precision, (c) Recall, (d) f1-score.

**Table 7**

RMSE comparison among FHMM_EXACT, CO, Mean, SIQCP and eFHMM-TS on synD dataset.

|  | CO | FHMM EXACT | Mean | SIQCP | eFHMM-TS |
|---|---|---|---|---|---|
| Fridge | 15.6 | 16.9 | 23.1 | 53.49 | 18.42 |
| Kettle | 42.1 | 34.3 | 44.1 | 80.13 | 21.93 |
| Toaster | 25.5 | 33.9 | 52.9 | 87.01 | 1.15 |
| Hair Dryer | 45.7 | 41.5 | 74.1 | 123.47 | 12.08 |
| Coffee Maker | 33.2 | 14.9 | 51.4 | 65.47 | 17.45 |
| Fan | 16.9 | 6.9 | 17.2 | 16.0 | 16.46 |
| Electric Space Heater | 26.9 | 28.3 | 198.7 | 31.3 | 10.58 |

Table 7 lists the RMSE of different methods on synD. As analyzed in this section and shown in Fig. 15, SIQCP has the highest RMSE and the worst performance. It proves that this kind of methods cannot fully exploit the advantages of high-resolution dataset on load disaggregation, and they may actually lead to worse performance on higher resolution dataset.

### 5.7 Comparison of computational time

As discussed before, the proposed event-driven NILM could disaggregate composite load in real-time no matter how high the sampling rate is. The computational time is proportional to the event number for a given household because it does not perform point-to-point inference compared with nonevent-based method. Some of the nonevent-based methods such as SIQCP in [18] segment the aggregated power and use a batch of data to infer appliance states that is more efficient than point-to-point inference. However, the number of segments increases exponentially when appliances with high fluctuation contribute to power consumption. It is the reason why event detection is essential to event-driven NILM method. Even if the number of appliances increases, the operation times of appliances per day is still limited, so the computational time is usually limited. FHMM_EXACT highly depends on the preset parameters, i.e., "*sample rate*" and "*chunck size*", which are set to 50 and 1000 in this paper for the purpose of comparing computational time.

As reported in [18], the total time complexity of SIQCP could be about $O(TK^N)$ in theory, which increases exponentially with the number of appliances. As discussed before, the computational complexity of eFHMM-TS is $O(MNK)$ that is linear to the appliance number. The time complexity of FHMM_EXACT depends on the two preset parameters and the appliances number.

Fig. 16 shows the comparison of computational time for SIQCP, FHMM_EXACT and eFHMM-TS on 50 Hz LIFTED. SIQCP_5 means that the threshold of SIQCP is 5watt and SIQCP_15 refers to setting threshold to 15watt. Smaller threshold would generate more segments. The tested data includes 1,951,235 datapoints for 650.41 minutes. eFHMM-TS could disaggregate composite load of 10 appliances in 68.40 seconds. FHMM_EXACT takes about 407 seconds. It can be observed that as the number of appliances increases, computation time of FHMM_EXACT and eFHMM-TS increases linearly as shown in the inner plot of Fig. 16. However, the computational times of SIQCP_5 and SIQCP_15 increase much higher on the same test data.

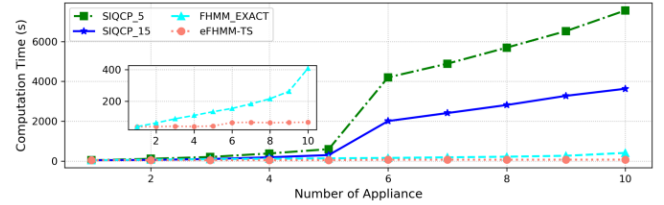

Fig. 16 Computational time comparison.

When the number of appliances increases to 6, the computational times for SIQCP and eFHMM-TS increase significantly. That is because the sixth appliance is washing machine that frequently turns on and off in the wash mode. As a result, the event number of washing machine is far more than other appliances. For SIQCP, the number of segments increases accordingly as well.

### 5.8 Size reduction of transmitted data using edge-cloud framework

One challenge of the proposed two-stage NILM solution is that its use of high-resolution data puts a lot of burdens on data storage and communication. To effectively overcome this challenge, the proposed eFHMM-TS solution method is integrated into the edge-cloud framework as shown in Fig. 7. Event detection is implemented on the edge side close to the resource-poor IoT devices such as smart meters and e-monitors. Transient signatures such as DTS and DSP and steady-state signatures such as SSP of each sliding window are transmitted to the cloud server.

One task of cloud server is to train models for each household and distribute the target appliance model if necessary. Thus, the extracted information from each appliance will also be transmitted to cloud server. If eFHMM-TS is done only when an event occurs, only two transient signatures and one steady signature will be transmitted to cloud server, the size of transmitted data as a percentage of the size of the raw measurement data (the percentage for short) is associated with event number $M$, i.e., $\frac{3M}{T} \times 100\%$. For the 50 Hz LIFTED, there are 488 events on 1,051,106 test datapoints, so the percentage is about 0.14%.

If the extracted SSP of each window is transmitted for the second stage confirmation, the percentage for each appliance is about $\frac{1}{n_w*f} \times 100\%$, where $n_w$ refers to $n$ seconds of window length by default, $f$ is the data sampling rate. For the test data in Fig. 8, $n_w$ are 2 seconds and $f$ is 50 Hz. So, as shown in Fig. 17, the percentages for all appliances are all around 1%.

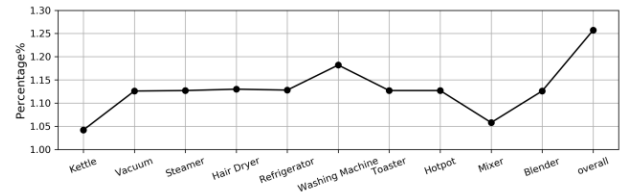

Fig. 17 The size of transmitted data as a percentage of the size of the raw measurement data using edge-cloud framework.

Furthermore, according to the characteristics of an appliance, the percentage is slightly different due to its event number and power fluctuation. As an example, washing machine frequently turns on and off in the wash mode to simulate hand wash, which leads to a greater number of events and associated transient signatures. Power fluctuation will modify the window length, i.e., $n_w$, thus changing the reduction percentage. So, the percentage for washing machine is the highest among all individual appliances.

## 6. Conclusions

High-resolution load measurement time series contains rich information to help disaggregate the load thus leading to a high-performance NILM solution. However, existing methods have not paid enough attention to the transient load signatures and usually suffer from high computational complexity when using high-resolution data. This paper proposes an event-driven NILM solution to solve this problem. It studies the characteristics of each appliance and their components and finds out the significance of transient signatures for appliance identification, which inspires us to develop the event-driven factorial HMM (eFHMM) for a household and apply the sparse Markov chains to realize online disaggregation.

A two-stage NILM solution is presented as well to infer state-changing appliance in the first stage and confirm the results of first stage using steady-state information in the second stage. To make the proposed solution a success in practice, we integrate the eFHMM-TS into edge-cloud framework to offload the storage and computation burden of IoT devices. The test results show that the proposed eFHMM-TS method enables highly accurate online appliance identification. The size of transmitted data is also reduced greatly as well.